\documentclass[12pt, preprint]{aastex}

\slugcomment{Accepted for Publication in {\it The Astrophysical Journal}}

\shorttitle{Pair cascades in IR radiation fields}
\shortauthors{P. Roustazadeh \& M. B\"ottcher}

\begin{document}

\title{Synchrotron Emission from VHE Gamma-Ray Induced Pair Cascades in
AGN Environments}

\author{P. Roustazadeh and M. B\"ottcher\altaffilmark{1}}

\altaffiltext{1}{Astrophysical Institute, Department of Physics and Astronomy, \\
Ohio University, Athens, OH 45701, USA}

\begin{abstract}
The discovery of very-high-energy (VHE, $E > 100$~GeV) $\gamma$-ray
emission from intermediate- and low-frequency peaked blazars suggests
that $\gamma\gamma$ absorption and pair cascading might occur in those
objects. In previous papers, we investigated the 
Compton emission from VHE $\gamma$-ray induced pair cascades, deflected 
by moderate magnetic fields, in a largely model-independent way, and 
demonstrated that this emission can explain the Fermi fluxes and spectra 
of the radio galaxies Cen~A and NGC~1275. In this paper, we describe a 
generalization of our Monte-Carlo cascade code to include the 
angle-dependent synchrotron output from the cascades, allowing 
for the application to situations with non-negligible magnetic fields, 
leading to potentially observable synchrotron signatures, but still 
not dominating the radiative energy loss of cascade particles.
We confirm that the synchrotron radiation from the cascades in NGC~1275 
and Cen~A are negligible for the parameters used in our previous works. 
We demonstrate that the magnetic field can not be determined from a fit
of the cascade emission to the $\gamma$-ray spectrum alone, and the
degeneracy can only be lifted if the synchrotron emission from the
cascades is observed as well. We illustrate this fact with the example 
of NGC~1275. We point out that the cascade synchrotron
emission may produce spectral features in the same energy range 
in which the big blue bump is observed in the spectral energy 
distributions of several blazars, and may make a non-negligible 
contribution to this feature. We illustrate this idea with the 
example of 3C~279.

\end{abstract}

\section{Introduction}

Among AGNs, blazars are the most luminous and violent objects in the
universe and emit $\gamma$-rays at energies higher than 100 MeV. They are
divided into two main subgroups: Highly
variable quasars, sometimes called optically violent variable (OVV) quasars,
and BL Lacertae objects (BL Lac). Since one of the jets of blazars is
pointed toward the Earth, we see the jet emission strongly Doppler enhanced and
highly variable. According to the unification scheme \citep{up95}, radio
galaxies are the mis-aligned parent population of blazars. Synchrotron peak
frequencies of BL Lac objects cover a large range from IR to X-rays, and based
on its location they are called low-frequency peaked BL Lac objects (LBL;
synchrotron peak in the IR), intermediate BL Lac objects (IBL; synchrotron
peak in the  optical/UV), or high-frequency peaked BL Lac objects (HBL;
synchrotron peak in the X-rays).

While there is little evidence for dense radiation environments
in the nuclear regions of BL~Lac objects --- in particular, HBLs
---, strong line emission in Flat Spectrum Radio Quasars (FSRQs)
as well as the occasional detection of emission lines in the spectra of some
BL~Lac objects \citep[e.g.,][]{vermeulen95} indicates dense nuclear
radiation fields in those objects. This is supported by spectral modeling
of the SEDs of blazars using leptonic models which prefer scenarios
based on external radiation fields as sources for Compton scattering
to produce the high-energy radiation in FSRQs, LBLs and also some
IBLs \citep[e.g.,][]{ghisellini98,madejski99,bb00,acciari08,abdo11}. If the
VHE $\gamma$-ray emission is indeed produced in the high-radiation-density
environment of the broad line region (BLR) and/or the dust torus of an
AGN, it is expected to be strongly attenuated by $\gamma\gamma$ pair
production \citep[e.g.][]{pb97,donea03,reimer07,liu08,sb08}.
 \cite{akc08} have suggested that such intrinsic $\gamma\gamma$ absorption may be
responsible for producing the unexpectedly hard intrinsic (i.e., after
correction for $\gamma\gamma$ absorption by the extragalactic background
light) VHE $\gamma$-ray spectra of some blazars at relatively high redshift.
A similar effect has been invoked by \cite{ps10} to explain the spectral
breaks in the {\it Fermi} spectra of $\gamma$-ray blazars.
This absorption process will lead to the development of Compton-supported
pair cascades in the circumnuclear environment \citep[e.g.,][]{bk95,sb10,rb10,rb11}.

In \cite{rb10,rb11}, we considered the full 3-dimensional development of
Compton-supported VHE $\gamma$-ray induced cascades in the external radiation
fields in AGN environments.
 In those works, we have left the origin (leptonic SSC or IC, or hadronic) 
of the primary VHE $\gamma$-ray emission deliberately unspecified in order to 
investigate the cascade development in as model-independent a way as possible.
We have shown that even very weak magnetic fields $(B\lesssim
\mu$G) may be sufficient for efficient quasi-isotropization of the cascade emission.
We applied this idea to fit the {\it Fermi}
$\gamma$-ray emission of the radio galaxies NGC~1275 and Cen~A. In
\cite{rb10,rb11}, parameters were chosen such that the synchrotron emission
from the cascades was negligible.

In this paper, we present a generalization of the Monte-Carlo cascade
code developed in \cite{rb11} to non-negligible magnetic fields and consider
the angle dependent synchrotron emission from the cascades. In section
\ref{setup}, we will outline the general model setup and assumptions
and describe the modified Monte-Carlo code. Numerical results for generic
parameters will be presented in section \ref{parameterstudy}. We confirm
that for the objects Cen~A and NGC~1275 the synchrotron radiation from the
cascades is negligible for those parameters used in \cite{rb10,rb11}. In section
\ref{degeneracy}, we investigate the effect of the magnetic field and its
degeneracy. We show that by studying only the high energy emission from
the cascades, the magnetic field can not be determined, and additional
constraints are needed from the synchrotron emission component. This
is illustrated for the case of NGC~1275. In section \ref{3C279}, we will
demonstrate that for moderately strong magnetic fields the synchrotron
emission from the cascades can produce a signature resembling the big blue
bump (BBB) observed in several blazars and demonstrate 
that this may make a non-negligible contribution to UV -- soft X-ray
SED. We illustrate this for the case of 3C~279. We summarize in Section 
\ref{summary}.

\section{\label{setup}Model Setup and Code Description}

The general model setup used for this work is described in \cite{rb10,rb11}.
The primary VHE $\gamma$-ray emission is represented as a mono-directional
beam of $\gamma$-rays propagating along the X axis, described by a power-law 
with photon spectral index $\alpha$ and a high-energy cut-off at $E_{\gamma, max}$. 
We assume that the primary $\gamma$-rays interact via $\gamma\gamma$
absorption and pair production with an isotropic radiation field with 
arbitrary spectrum within a fixed boundary, given by a radius $R_{\rm ext}$.

\begin{figure}[ht]
\vskip 1.5cm
\centering
\includegraphics[width=12cm]{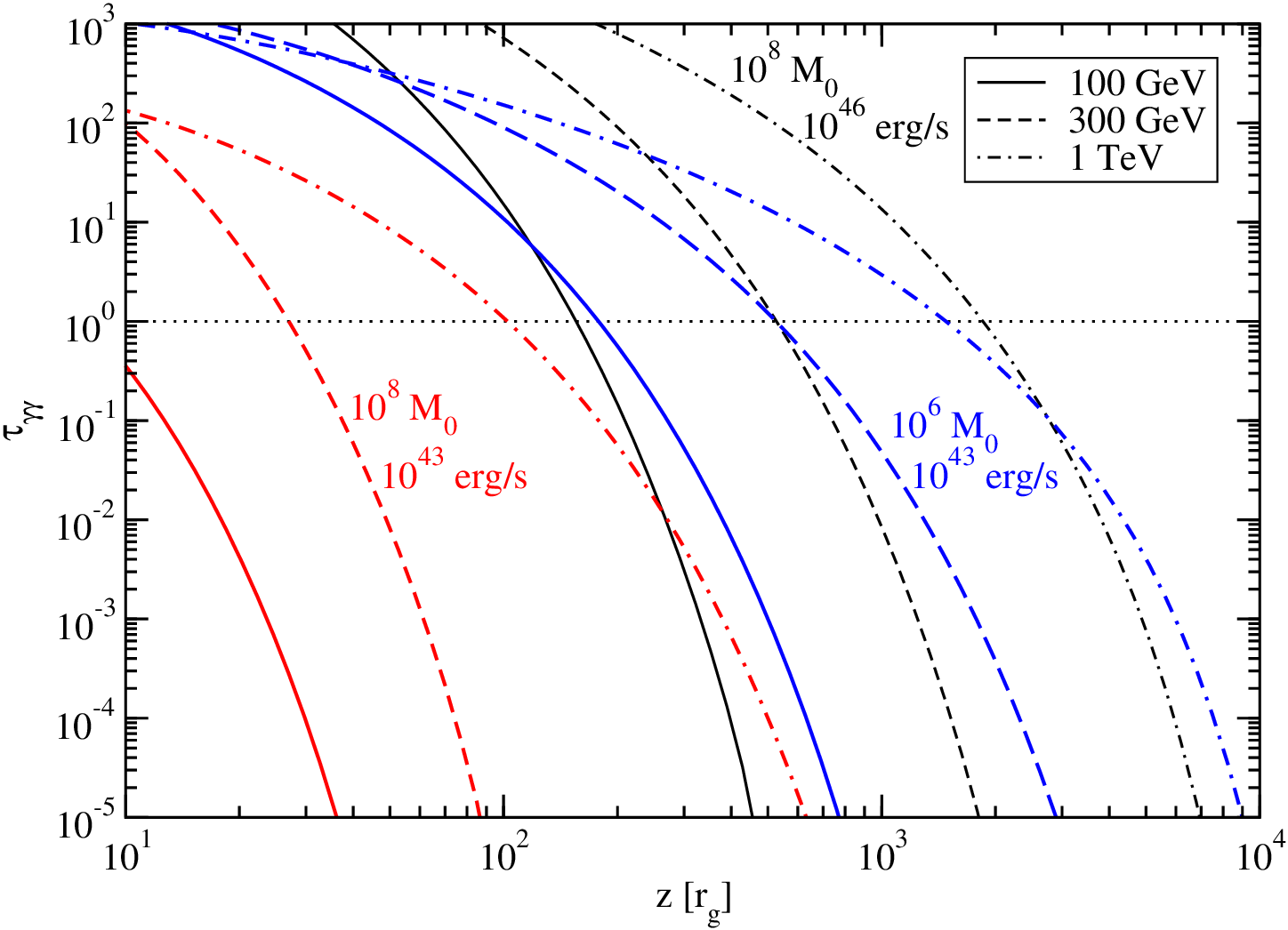}
\caption{\label{diskabs}$\gamma\gamma$ opacity due to accretion
disk photons as a function of height $z$ (in units of gravitational
radii $r_g = GM/c^2$) of the emission region above the black hole,
for three different VHE $\gamma$-ray photon energies, and different
black-hole masses and disk luminosities. }
\end{figure}

\begin{figure}[ht]
\vskip 1.5cm
\centering
\includegraphics[width=12cm]{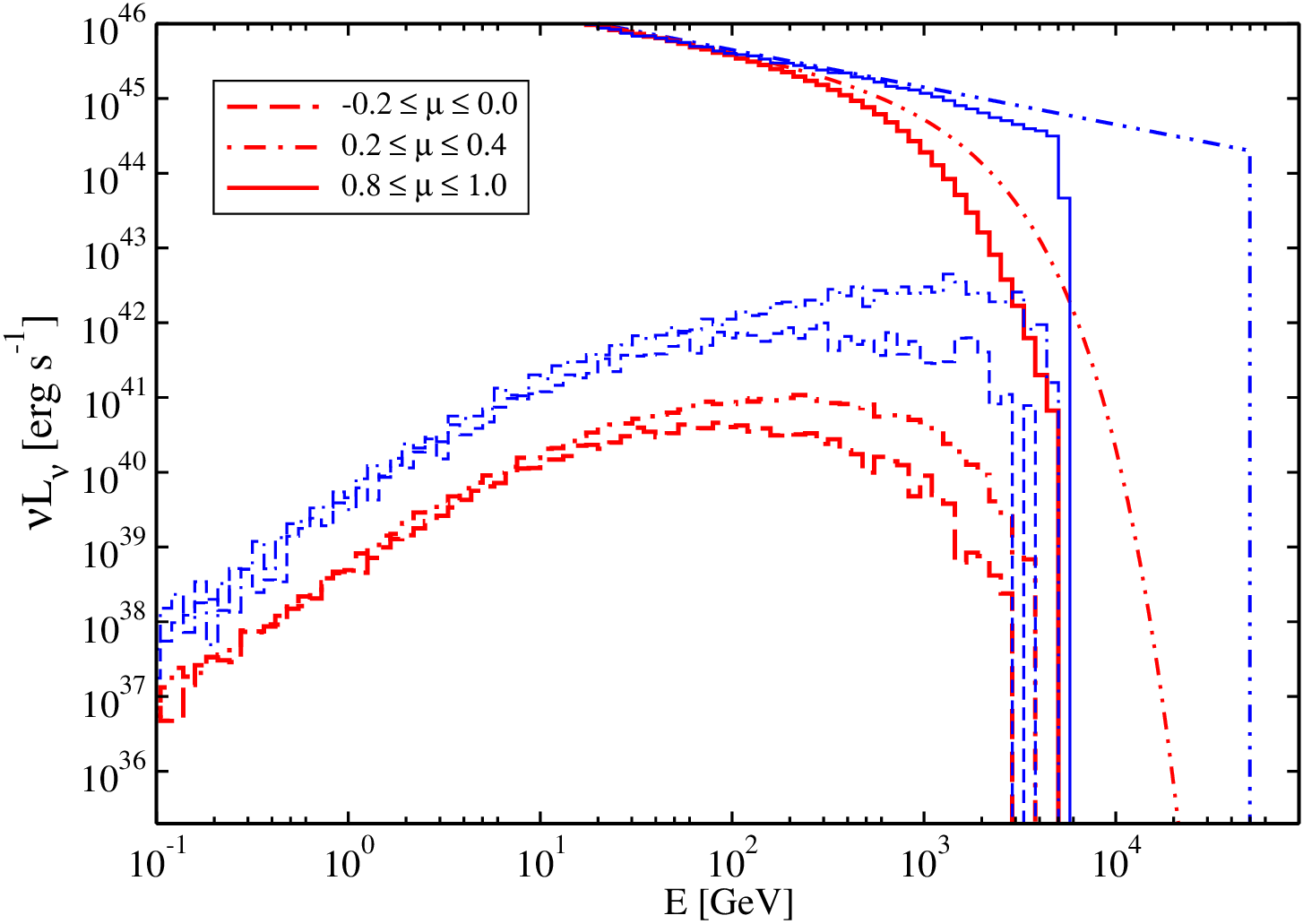}
\caption{\label{PLExpcomparison}Comparison of the Compton emission
from cascades for two different input VHE $\gamma$-ray spectral
shapes: Blue (thin) lines indicate the emission for a pure power-law
input spectrum; red (thick) lines for a power-law + exponential cut-off.
Environmental parameters are the same as in Figure \ref{standardfig}
(see below). Different line styles correspond to different viewing
angles, $\mu = \cos\theta_{\rm obs}$, with respect to the jet axis,
as indicated in the legend. Parameters: $B_x = B_y = 1$~$\mu$G, 
$\theta_B = 45^o$; $u_{\rm ext} = 10^{-6}$~erg~cm$^{-3}$, 
$R_{\rm ext} = 10^{18}$~cm, $T = 1000$~K. 
The angular bin $0.8 \le \mu \le 1$ contains the forward direction. 
The (unabsorbed) primary $\gamma$-ray input spectra are shown by the 
dot-dot-dashed lines.
}
\end{figure}

The assumption of an isotropic external radiation field is appropriate
for line emission from the BLR, for distances from the central engine comparable
to the size of the BLR ($\sim 10^{17}$ -- $10^{18}$~cm), and for infrared emission
from cold dust in the nuclear environment, on typical scales of $\sim$~parsec.
Close to the central black hole and accretion disk, direct emission from the
accretion disk may dominate the radiation energy density. However, for moderate
distances from the disk, the primary $\gamma$-ray beam will interact with the
accretion disk emission under an unfavorable angle for $\gamma\gamma$ pair
production. To illustrate this point, we plot in Figure \ref{diskabs} the opacity
to $\gamma\gamma$ absorption in the radiation field of an optically thick,
geometrically thin \citep{ss73} accretion disk as a function of height
$z$ of the emission region from the black hole. The figure shows that
typically, the accretion-disk $\gamma\gamma$ opacity drops below one
at distances of a few 100 -- $10^3 \, r_g$ from the black hole. This
is of the order of the characteristic height of the emission region in
leptonic models of blazar emission, and much smaller than the size of
the BLR or the dust torus. We therefore conclude that the neglect of
the accretion-disk emission in our simulations is a good approximation
throughout almost all of our simulation volume.
For the case of thermal blackbody radiation fields considered below, we
choose energy densities and blackbody temperatures characteristic of the
observed properties (temperatures and total luminosities) of thermal 
infrared emission seen in AGN.
In order to keep our treatment as model independent as possible, we do not 
specify the physical origin of the primary VHE $\gamma$-ray spectrum. The
simplest assumption consistent with most models of $\gamma$-ray emission
in blazars is a simple power-law, which we use as input spectrum in our
simulations. The shape of the resulting pair cascade emission is only 
very weakly dependent on the exact shape of the incident VHE $\gamma$-ray 
spectrum. This is illustrated in Figure \ref{PLExpcomparison}, in which
we run a cascade simulation, once with a straight power-law, once with
a powerlaw + exponential cut-off (as a more realistic representation of
a physical blazar high-energy spectrum), with identical environmental
parameters, as listed in the figure caption. While the overall normalization
of the cascade spectrum, obviously, depends on the flux of absorbed
VHE $\gamma$-rays (which is higher in the pure power-law case), the
cascade spectra at energies below the $\gamma\gamma$ absorption
trough are virtually identical. In the cases relevant for this study,
only a small fraction of the $\gamma$-ray power is absorbed and re-processed
into cascades. Therefore, feedback between the primary $\gamma$-ray production
region and the cascade emission may be neglected, and the cascade development
can be treated as a process separate from the (unspecified) VHE $\gamma$-ray
production mechanism.

Our code evaluates $\gamma\gamma$ absorption and pair production using the
full analytical solution to the pair production spectrum of \cite{bs97} under
the assumption that the produced electron and positron travel initially along
the direction
of propagation of the incoming $\gamma$-ray. The trajectories of the particles are
followed in full 3-D geometry. Compton scattering is evaluated using the head-on
approximation, assuming that the scattered photon travels along the direction of
motion of the electron/positron at the time of scattering. The Compton
energy loss to the electron is properly accounted for at the time of each scattering.

For simplicity, the magnetic field in our simulations is treated as 
homogeneous, oriented at an angle $\theta_B$ with respect to the jet axis.
This may be considered an appropriate proxy for a helical magnetic field
with the ratio of toroidal ($B_{\rm tor}$) and poloidal ($B_x$) magnetic
fields given by $\tan\theta_B = B_{\rm tor}/B_x$.
The code calculates the synchrotron energy loss of cascade particles in the
following way: The energy of electrons/positrons is decreased by $\Delta E_{\rm sy}
= \dot E_{\rm sy} \frac{l_c}{c}$ between successive Compton scatterings, where
$l_c$ is the Monte Carlo generated distance traveled to the next scattering
and $ \dot E_{\rm sy} = - 2 c \sigma_{T} u_B \gamma^2 \sin^2 \psi$ with
$\psi$ being a pitch angle between the particle momentum and the magnetic field.
We assume that the trajectory of the particles between two Compton scatterings
is not affected by synchrotron radiation which is valid for $ u_B \lesssim u_{\rm ext}$.
Then for $10$ random points between two successive Compton scatterings, we determine
the position and direction of motion of the particles at these points and write
the spectral power in synchrotron radiation $P_{\nu}$ into a synchrotron output
file for the angular bin corresponding to the electron's/positron's direction of
motion. The synchrotron power $P_{\nu}$ of a single $e^{\pm}$ is approximated as:

\begin{equation}
P_{\nu} = 2 \frac{c \sigma_T}{\Gamma(\frac{4}{3})} u_B \beta^2\gamma^2
\frac{\nu^{1/3}}{\nu_c^{4/3}} e^{-\nu/\nu_c}
\label{sy_asymptotic}
\end{equation}

\citep{boettcher12} where the critical frequency 
$\nu_c = \frac{3 q B}{4\pi m_e c}\gamma^2 \sin\psi = 
4.2\times10^6 \sin\psi B_G \gamma^2$~Hz with $B_G$ being the magnetic 
field in units of Gauss. 
The approximation (\ref{sy_asymptotic}) represents the synchrotron
spectrum to whithin a few \% at all frequencies.

\section{\label{parameterstudy}Numerical Results}

\begin{figure}[ht]
\vskip 1.5cm
\centering
\includegraphics[width=12cm]{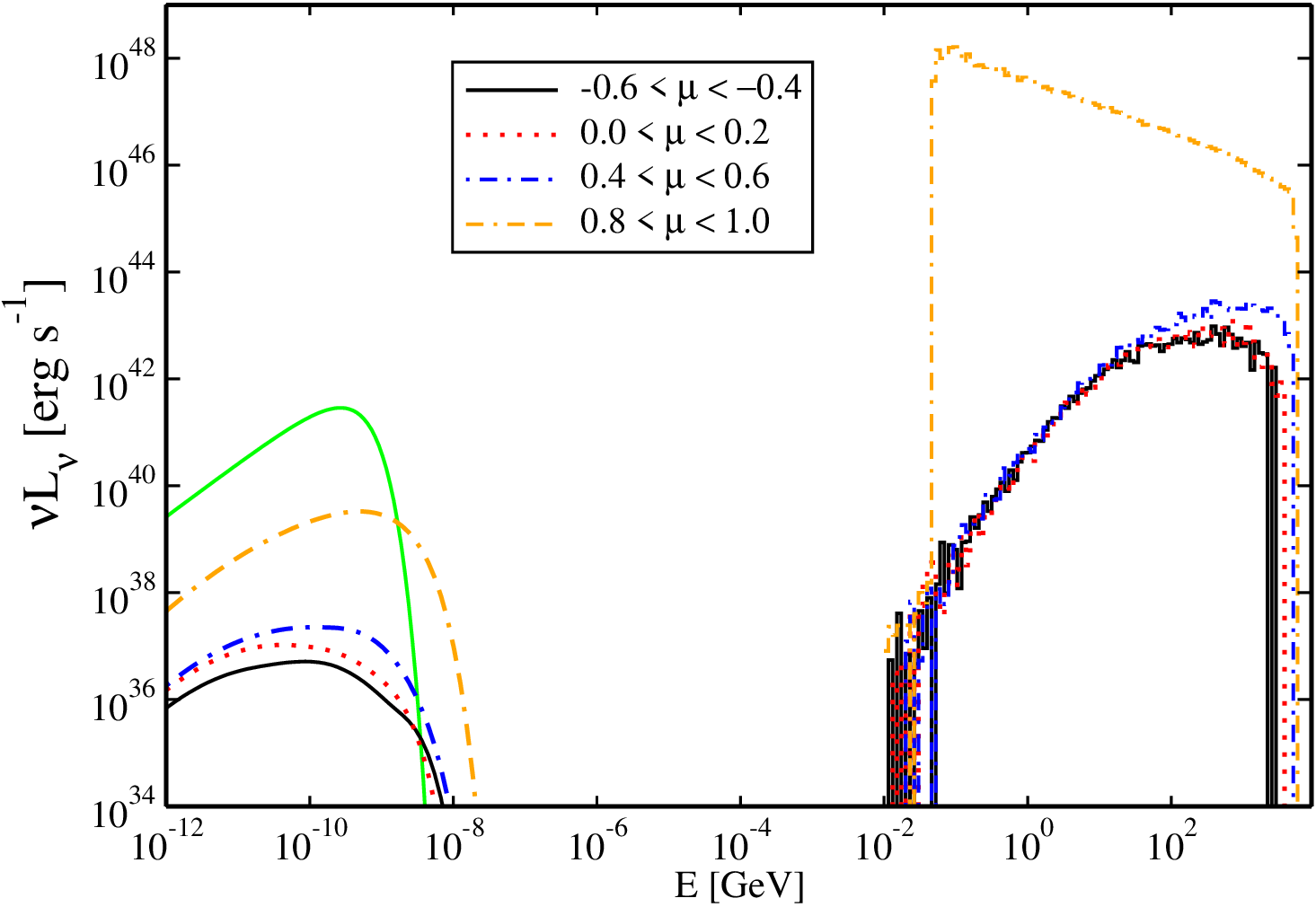}
\caption{\label{standardfig}
Cascade emission at different viewing angles ($\mu = \cos\theta_{\rm obs}$). 
Parameters of the target photon field are the same as for Figure 1. The 
input photon spectrum is a pure power-law with $\alpha = 2.5$, 
{$E_{\gamma, {\rm max}} = 5$~TeV}. The green solid line represents 
the target photon field.
}
\end{figure}

We have used the cascade Monte-Carlo code described in the previous section
to evaluate the angle-dependent Compton and synchrotron spectra from VHE
$\gamma$-ray induced pair cascades for a variety of generic
parameter choices. Figure \ref{standardfig} illustrates the viewing angle
dependence of the cascade emission. For this simulation, we assumed a
magnetic field of $B = \sqrt{2} \mu$G, oriented at an angle $\theta_B = 45^o$
with respect to the X axis ($B_x = 1 \, \mu$G, $B_y = 1 \, \mu$G). The
external radiation field is a thermal blackbody with $u_{\rm ext} =
10^{-6}$~erg~cm$^{-3}$, extended over a region of radius $R_{\rm ext} = 
10^{18}$~cm, with a blackbody temperature of $ T = 10^3$~K (corresponding 
to a peak of the blackbody spectrum at a photon energy of $E_s^{\rm pk}= 
0.25$~eV). This leads to a $\gamma\gamma$ absorption cut-off at an energy
$E_c = (m_e c^2)^2 / E_s \sim 2$~TeV. The incident $\gamma$-ray spectrum 
has a photon index of $\alpha = 2.5$ and extends out to $E_{\gamma, {\rm max}} 
= 5$~TeV.

For any given viewing angle $\theta$ with respect to the direction of 
propagation of the primary $\gamma$-rays a critical electron energy for 
which the deflection angle over a Compton length equals the observing 
angle, i.e., $\theta \sim \lambda_{\rm IC} / r_g$ can be defined. This 
yields the characteristic electron energy $E_{\rm e, br} = \gamma _c 
m_e c^2$ corresponding to a given observing angle $\theta$:

\begin{equation}
\gamma_c =\sqrt{\frac {3  e B}{4 \sigma_T  u_{\rm ext}\theta}}\sim 7.2\times10^5
B_{-6}^{1/2} u_{-3}^{-1/2} \theta^{-1/2}
\end{equation}
where $B_{-6} = B/\mu$G and $u_{-3} = u_{\rm ext}/(10^{-3}$~erg~cm$^{-3}$).

This expression has been derived assuming that the Compton cooling
length can be calculated in the Thomson regime, which is valid for
$\gamma \lesssim 2 \times 10^6 T_3^{-1}$ for a thermal target photon
field with temperature $T = 10^3 \, T_3$~K, or $\gamma \lesssim 5 \times 
10^4$ for a Ly$\alpha$-dominated target photon field.
If these electrons radiate their energy by synchrotron radiation and Compton
upscattering with the soft photon field in the Thomson regime, we can find the
corresponding spectral breaks for synchrotron radiation and Compton scattering
as a function of viewing angle:

\begin{equation}
E_{\rm sy,br}\cong \gamma_c^2 B m_e c^2/B_{cr}= \frac{ 3m_e c^2  e B^2}{4
\sigma_T  u_{\rm ext}
\, \theta \ B_{\rm cr}}\sim 6.15 B_{-6}^2 u_{-3}^{-1}
\theta^{-1}{\rm meV}
\label{ViewingangleSy}
\end{equation}

\begin{equation}
E_{\rm IC, br}\cong\gamma_c^2 E_s = {3 \, e \, B \over 4 \, \sigma_T \, u_{ext}
\, \theta}
\, E_s \sim 5.4\times 10^2 \, E_{s,1}B_{-6} u_{-3}^{-1} \theta^{-1}\; {\rm GeV}.
\label{ICbreak}
\end{equation}
where the $B_{\rm cr}= \frac{m_e^2 c^3}{e \hbar} = 4.4 \times10^{13}~G$ and
$E_{s,1} = E_s/(1$~eV).

Therefore, the ratio of the Compton to the synchrotron peak frequency is given by:
\begin{equation}
\frac{E_{\rm IC, br}}{ E_{\rm sy,br}} = \epsilon_s \frac{B_{\rm cr}}{B}
\label{RatioComToSyn}
\end{equation}
where $\epsilon_s = \frac{E_s}{m_e c^2}$.

Figure \ref{standardfig} shows that with increasing viewing angle, the
spectral peaks of both the synchrotron and Compton emission shift to
lower energy. This is because the Compton cooling length of the high
energy particles is much smaller than their Larmor radius $\lambda_{IC}\ll
r_g$, so they are emitting while traveling in the forward direction. Instead,
for low energy particles, $\lambda_{IC}\geq r_g$, so that they are deflected
before they are emitting. For the Compton emission this effect was already
discussed in \cite{rb10,rb11}.

\begin{figure}[ht]
\vskip 1.5cm
\centering
\includegraphics[width=12cm]{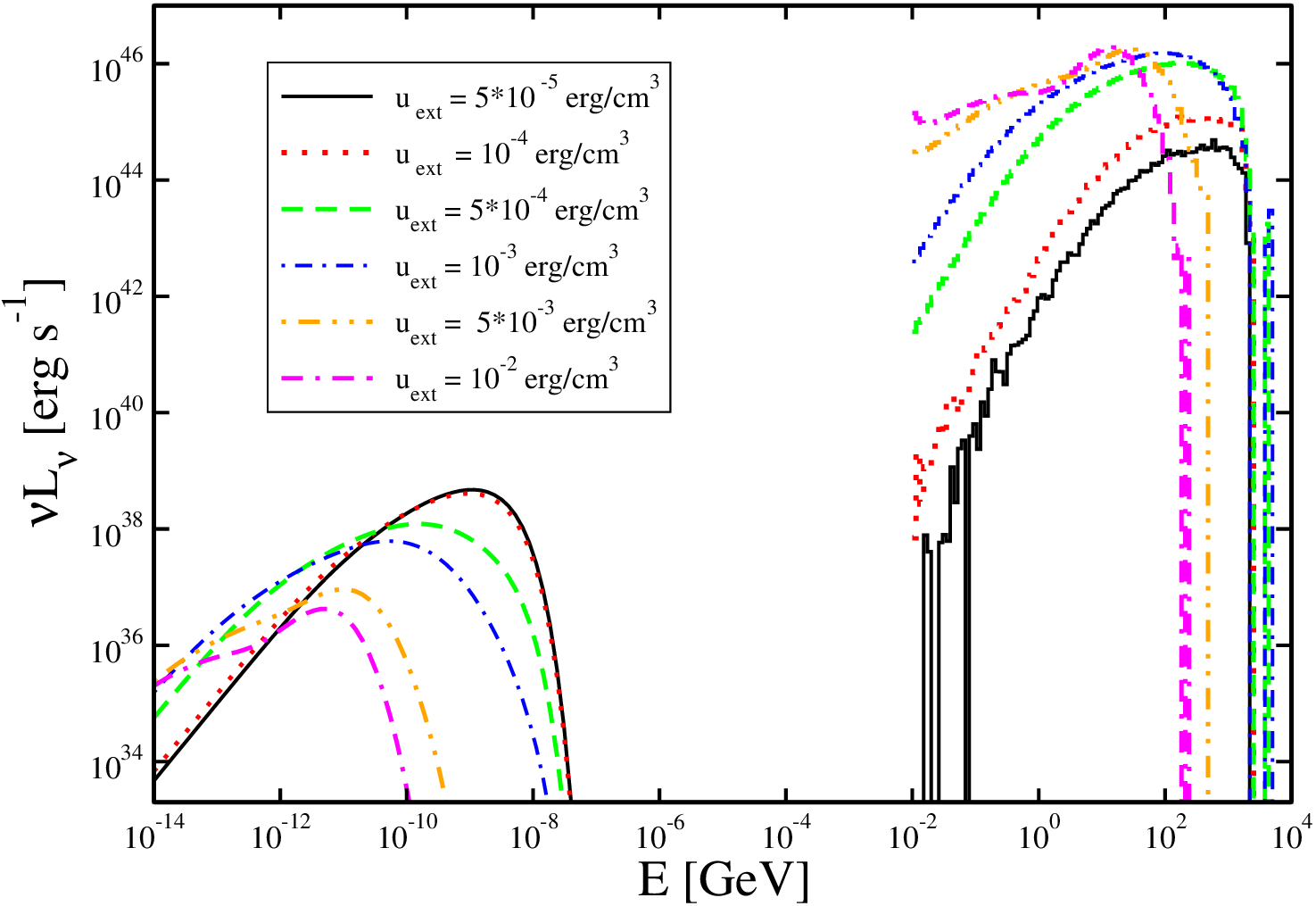}
\caption{\label{ufig}The effect of a varying external radiation energy density.
Parameters: $B_x = B_y = 10^{-6}$~G; $R_{\rm ext} = 10^{16}$~cm, $T = 10^3$~K;
$\alpha = 2.5$, $E_{\gamma, {\rm max}} = 5$~TeV. The cascade emission in the 
angular bin $0.2 \leq \mu \leq 0.4$ is shown.}
\end{figure}

Figure \ref{ufig} shows the cascade spectra for different values
of the external radiation field energy density $u_{\rm ext}$. In accordance
with equations \ref{ViewingangleSy} and \ref{ICbreak}, for higher values of
the external radiation field, the spectral breaks of both radiation components
shift to lower energies. Figure \ref{ufig} also shows that the synchrotron
luminosities of the cascades decrease with increasing $u_{\rm ext}$ while
the Compton luminosities of the cascades increase. For a larger value of
$u_{\rm ext}$ and fixed blackbody temperature the soft target photon number
density increases and $\tau_{\gamma\gamma}$ becomes larger so that the number
of VHE photons which will be absorbed increases and the photon flux of Compton
emission from the cascades becomes larger. For very large values of $u_{\rm ext}$,
$\tau_{\gamma\gamma}\gg 1$ for photons above the pair production threshold so
that essentially all VHE photons will be absorbed and the Compton flux from
the cascade becomes independent of $u_{\rm ext}$ \cite[]{rb10,rb11}. The ratio
of emitted power in Compton to synchrotron radiation in the linear regime
$(\tau_{\gamma\gamma} \lesssim 1)$ is given by:
\begin{equation}
\frac{P_{sy}}{P_{IC}}=\frac{B^2 / 8 \pi}{u_{ext}}
\label{RatioCom}
\end{equation}
if Compton scattering occurs in the Thomson regime.
The flux ratio $\frac{F_{sy}}{F_{IC}} \varpropto
u_{\rm ext}^{-1}$, so that by increasing the $u_{\rm ext}$ the synchrotron
flux decreases.

\begin{figure}[ht]
\vskip 1.5cm
\centering
\includegraphics[width=12cm]{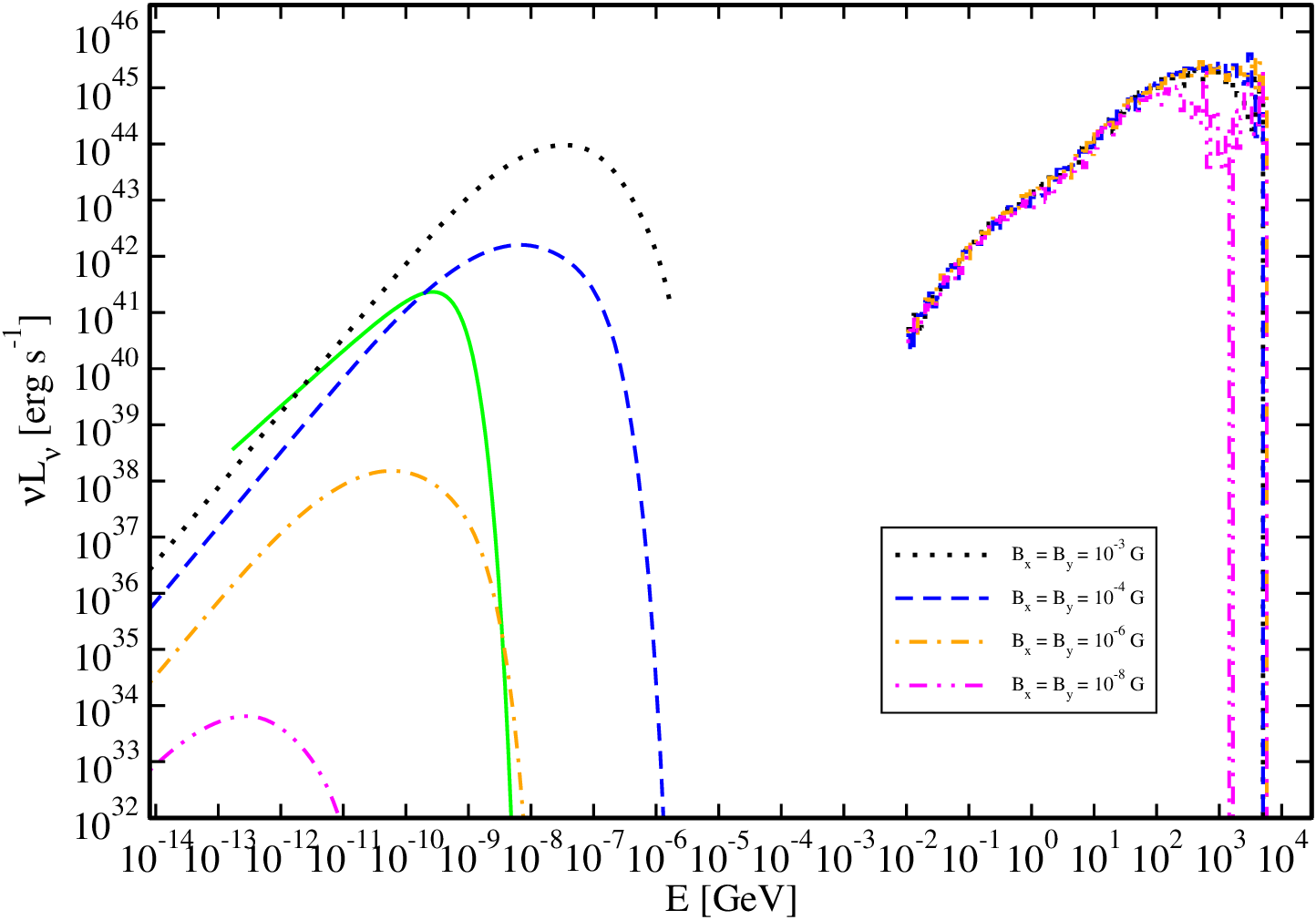}
\caption{\label{Bfig}The effect of a varying magnetic field strength for a
fixed angle of $\theta_B = 45^o$ between jet axis and magnetic field. Parameters:
$u_{\rm ext} = 10^{-6}$~erg~cm$^{-3}$, $R_{\rm ext} = 10^{18}$~cm, $T = 1000$~K,
$\alpha = 2.5$, {$E_{\gamma, {\rm max}} = 5$~TeV}. The cascade emission in the
angular bin $0.2\leq\mu\leq0.4$ is shown. The solid green line represents 
the target photon field.
}
\end{figure}

Figure \ref{Bfig} illustrates the effect of a varying magnetic field strength
for fixed magnetic field
orientation ($\theta_B = 45^o$). We see that the synchrotron peak energy
increases proportional to the square of the magnetic field strength as expected
from Eq. \ref{ViewingangleSy}. As already discussed in \cite{rb10,rb11} the
Compton energy break is proportional to the magnetic field strength. as long
as it occurs below the $\gamma\gamma$ absorption cut-off energy. The synchrotron
flux is proportional to the square of the magnetic field strength.
The flux ratio is $\frac{F_{sy}}{F_{IC}} \varpropto B^2$ until the fluxes become
comparable, at which point our treatment of synchrotron losses breaks down.

\begin{figure}[ht]
\vskip 1cm
   \centerline{
        \includegraphics[width=8cm,height=8cm]{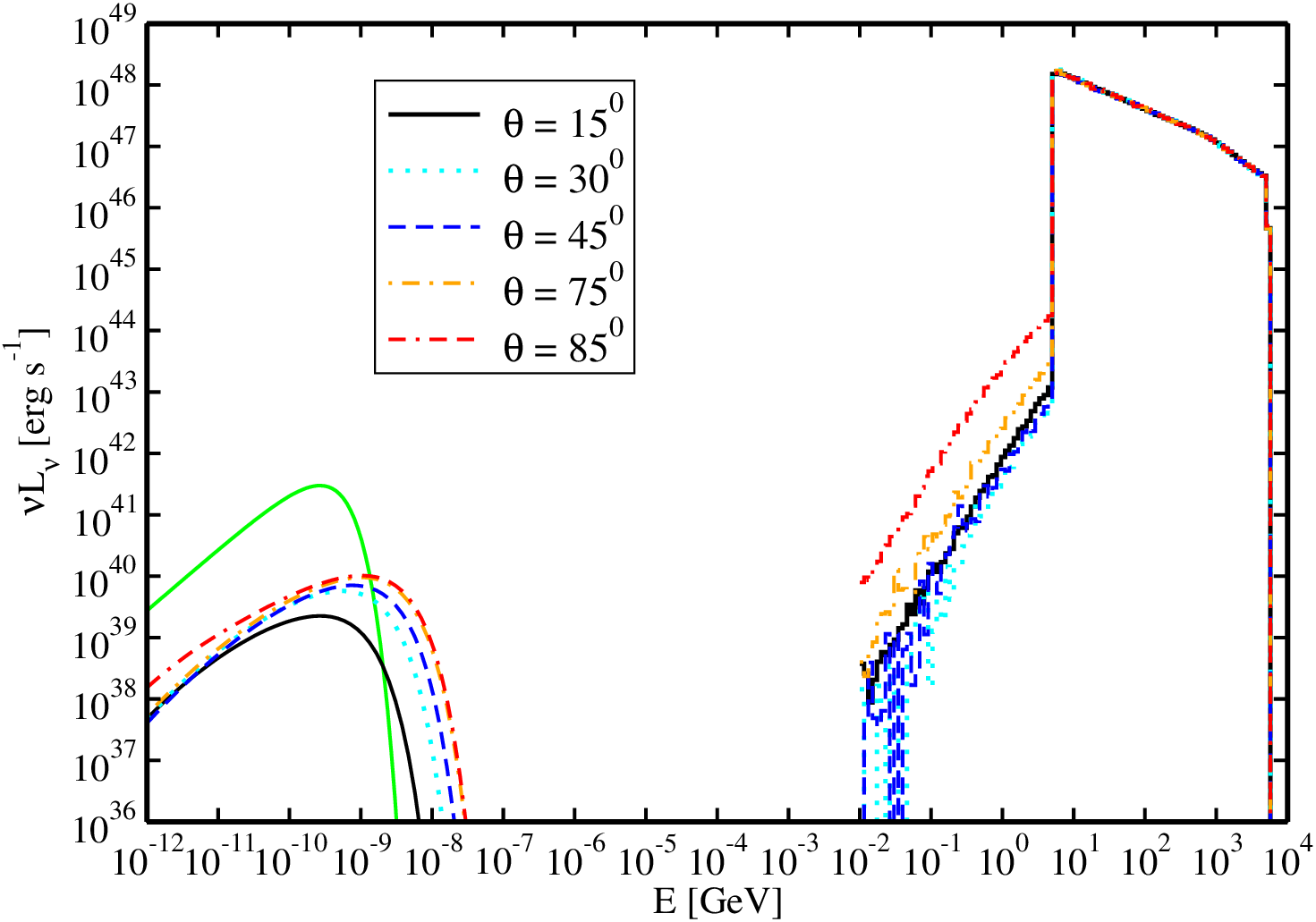}
        \includegraphics[width=8cm,height=8cm]{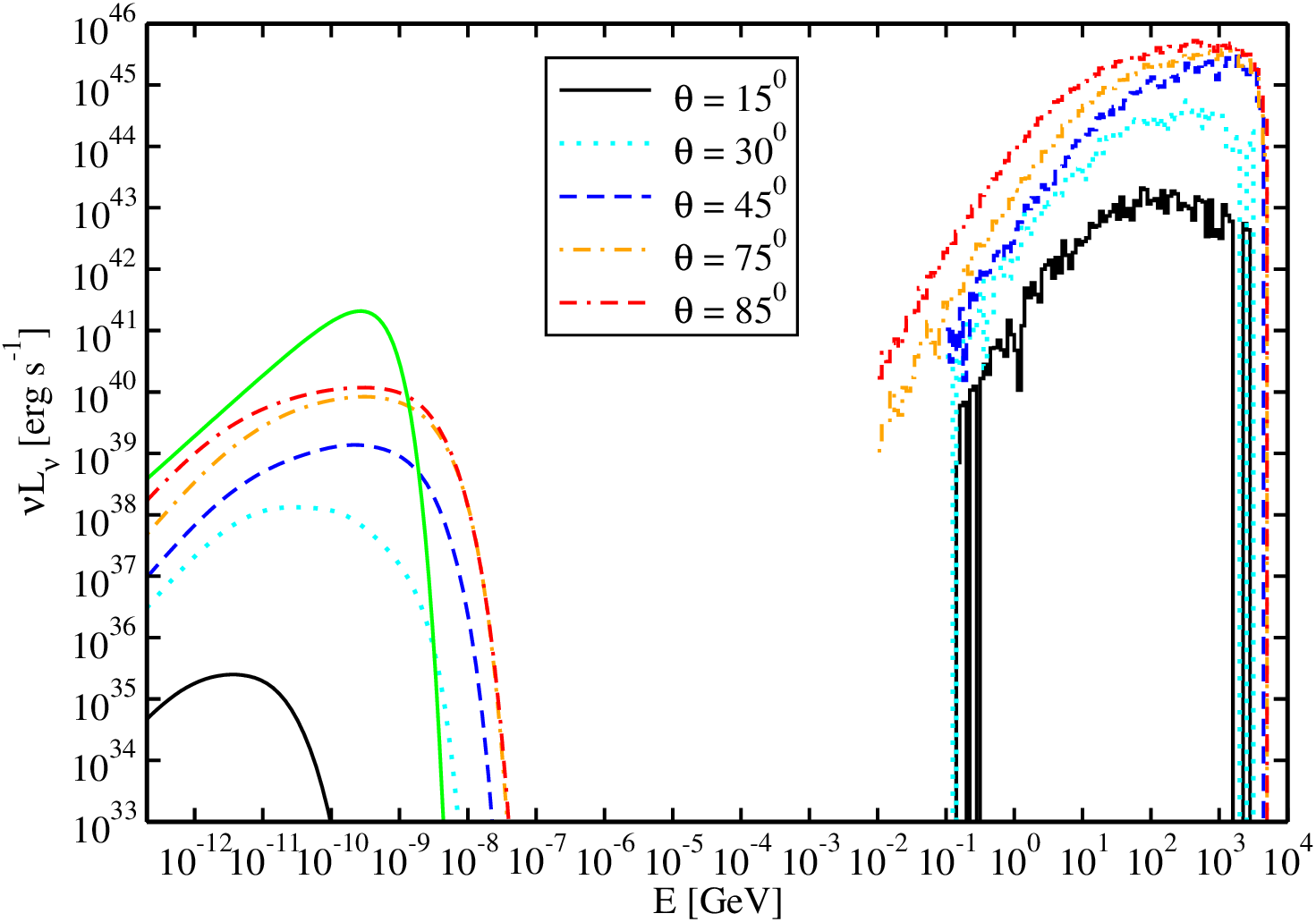}}
        \caption{\label{BfigF}The effect of a varying magnetic field orientation
for a fixed magnetic field strength of $B = 1 \, \mu$G, $u_{\rm ext} =
10^{-6}$~erg~cm$^{-3}$, $R_{\rm ext} = 10^{18}$~cm, $T = 1000$~K, $\alpha = 2.5$,
{$E_{\gamma, {\rm max}} = 5$~TeV}. Left figure: angular bin $0.8 \leq\mu\leq 1.0$
(dominated by the forward direction, i.e., the blazar case). Right figure: angular 
bin $0.2 \le \mu \le 0.4$, representative of radio galaxies.
The green solid lines represent the target photon fields.
}
    \end{figure}

Figure \ref{BfigF} illustrates the effects of a varying magnetic-field orientation
with respect to the jet axis, for fixed magnetic-field strength $B = 1 \, \mu$G
for different angular bins. The results for the Compton component have been
discussed \cite{rb11}. The figure illustrates that primarily the perpendicular
($B_y$) component of the magnetic field is responsible for synchrotron radiation.

\begin{figure}[ht]
\vskip 1cm
   \centerline{
        \includegraphics[width=8cm,height=8cm]{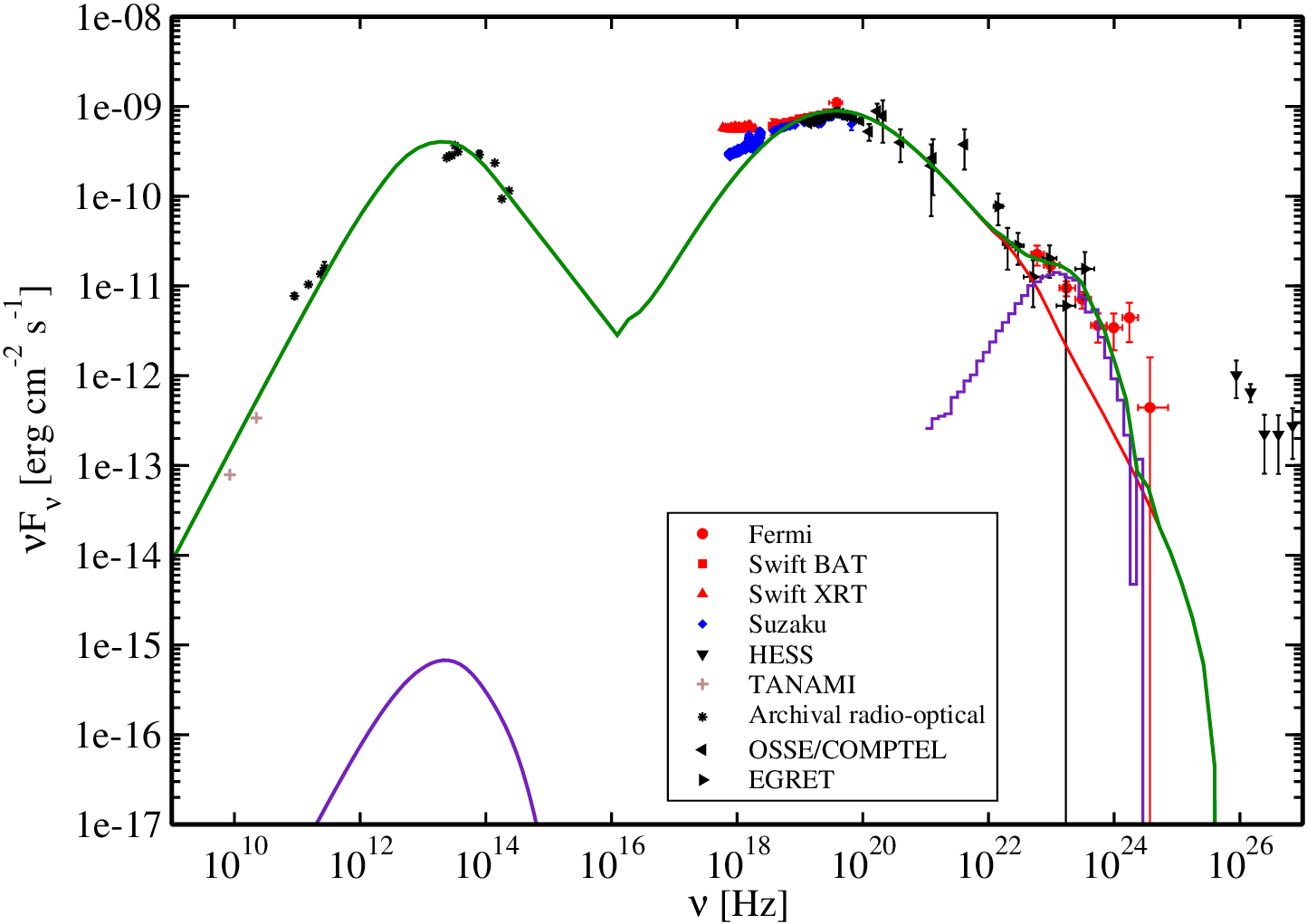}
        \hskip 1cm
        \includegraphics[width=8cm,height=8cm]{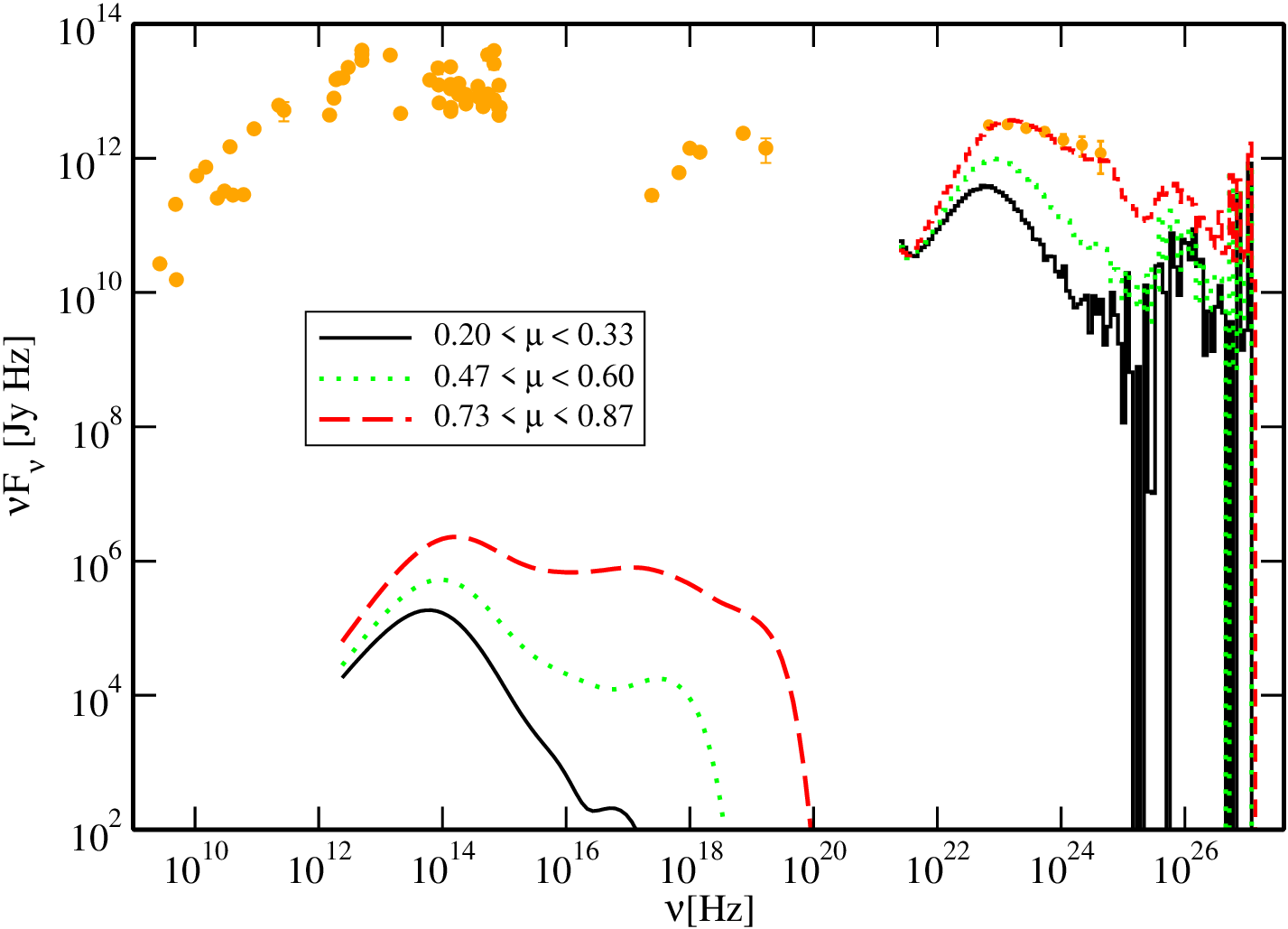}}
        \caption{\label{fitCenAandNGC1275}Compton and synchrotron radiation form
        the cascades. Left figure: (Fit to the SED of Cen~A.); Right figure:
        (spectrum of NGC~1275 with a simulated cascade spectrum from a mis-aligned
        blazar, along with the cascade spectra at larger viewing angles)}
     \end{figure}

Figures \ref{fitCenAandNGC1275} illustrate that the cascades emissions from
the synchrotron radiation for parameters used in \cite[]{rb10,rb11} are
negligible compared to the cascade Compton emission and much smaller than
the synchrotron radiation from the jet itself. This confirms that neglecting
synchrotron radiation in our previous works was justified.

\section{\label{degeneracy}Magnetic Field degeneracy}

\begin{figure}[ht]
\vskip 1.5cm
\centering
\includegraphics[width=12cm]{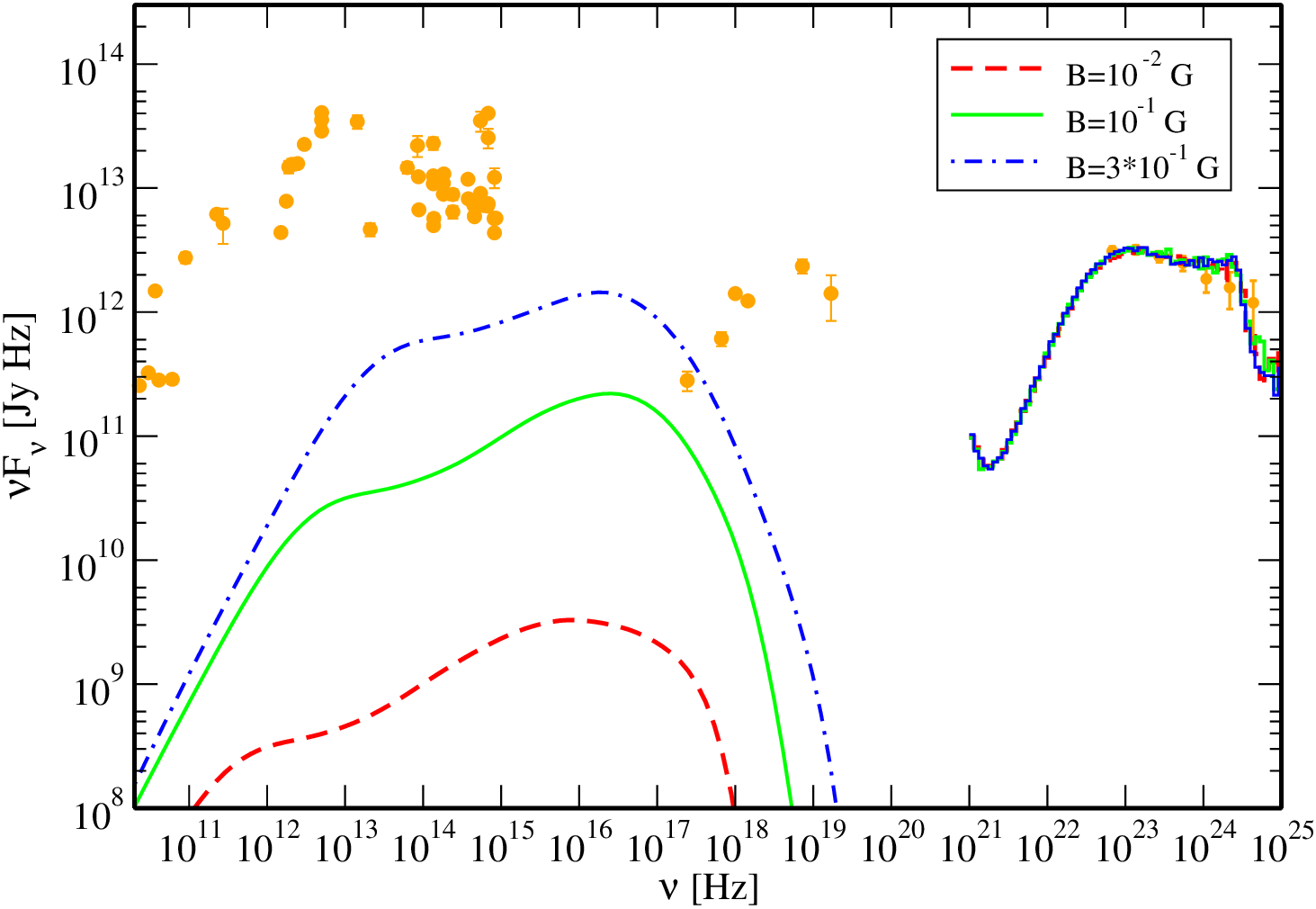}
\caption{\label{Degeneracy1}
Synchrotron and Compton emission form the cascades for NGC~1275
($0.6 \leq\mu\leq 0.8 $). Parameters: $\theta_B = 11^o$;
$u_{\rm ext} = 5\times10^{-2}$~erg~cm$^{-3}$, $R_{\rm ext} = 10^{16}$~cm,
$E_s=E_{L\alpha}$,
$\alpha = 2.5$, {$E_{\gamma, {\rm max}} = 5$~TeV}.
}
\end{figure}

In \cite{rb10}, we presented a fit to the \emph{Fermi} spectrum of the radio
galaxy NGC~1275. We now show that there is a degeneracy of the magnetic field,
both orientation and strength, if only the high energy output from the cascades
is considered. Figure \ref{Degeneracy1} shows this effect for NGC~1275. In this
plot, the external radiation field is
parameterized through $u_{\rm ext} = 5 \times 10^{-2}$~erg~cm$^{-3}$ with photon
energy $E_s = E_{Ly\alpha}$ and $R_{\rm ext} = 10^{16}$~cm. This size scale is
appropriate for low-luminosity AGN as observed in NGC~1275 \citep[e.g.][]{kaspi07},
and the parameters combine
to a BLR luminosity of $L_{\rm BLR} = 4 \pi R_{\rm ext}^2 \, c \, u_{\rm ext} =
1.9\times 10^{42}$~erg~s$^{-1}$, in agreement with the observed value for NGC~1275.
The magnetic field orientation is at an angle of $\theta_B = 11^o$. 
The mass of the black hole in NGC~1275 is uncertain, and estimates 
range from a few times $10^6 \, M_{\odot}$ \citep{levinson95} to 
$\sim 10^8 \, M_{\odot}$ \citep{wilman05}. Assuming a characteristic
fraction of $0.1$ of the accretion-disk luminosity to be re-processed
in the BLR, the accretion-disk luminosity may be estimated to be $L_D
\sim 10^{43}$~erg~s$^{-1}$. Figure \ref{diskabs} shows the $\gamma\gamma$
absorption depth due to the disk radiation field for $L_D = 10^{43}$~erg~s$^{-1}$
for the two possible extreme values of the black-hole mass, as a function
of height $z$ of the emission region above the accretion disk. It 
shows that for $M_{\rm BH} = 10^6 \, M_{\odot}$, the $\gamma\gamma$
opacity drops below one at $\sim 10^3 \, r_g \sim 10^{14}$~cm from 
the black hole, while for $M_{\rm BH} = 10^8 \, M_{\odot}$ $\gamma\gamma$
absorption becomes negligible at $\sim 10^2 \, r_g \sim \sim 10^{15}$~cm 
for primary $\gamma$-rays of $E_{\gamma} = 1$~TeV, and much earlier for
lower-energy photons. Therefore, throughout most of our simulation 
volume ($R_{\rm ext} = 10^{16}$~cm), $\gamma\gamma$ absorption in
the disk radiation field can be safely neglected.

The cascade spectrum shown in Figure \ref{Degeneracy1} pertains
to the angular bin $0.6 < \mu < 0.8$ (corresponding to $37^o \lesssim
\theta \lesssim 53^o$), appropriate for the known orientation of NGC~1275. In
\cite[]{rb10,rb11}, we have shown that for magnetic field values of $B\geq 1$~nG
and for energy density $u_{ext} \geq 10^{-3}$~erg~cm$^{-3}$, there is no pronounced
break in the cascade spectrum and the cascade is independent of magnetic field.
In general we expect no break in the cascade Compton emission if $E_{\rm IC,br}
\gtrsim \frac{(mc^2)^2}{E_s}$, which leads to the condition:

\begin{equation}
B \gtrsim \frac{{(m_e c^2)}^2 4\sigma_T u_{\rm ext} \theta}{3 e (E_s)^2} \sim 5 \,
u_{ext,-3} E_{s,1}^{-2} \theta \; {\rm nG}
\label{relation}
\end{equation}

Figure \ref{Degeneracy1} shows that while the high energy emission due to
deflection of the cascade up to the $\gamma\gamma$ absorption trough remains
the same for the different magnetic fields, the synchrotron emission from the
cascade changes. Therefore, determining the B field requires knowledge of the
synchrotron emission.

In the regime where $E_{\rm IC, br}$ is independent of the magnetic field,
$\nu_{\rm sy}\varpropto B$ according to Eq. \ref{RatioComToSyn} and the synchrotron
power is proportional to the square of magnetic field in agreement with figure
\ref{Degeneracy1}.

Since the synchrotron/Compton flux ratio $\frac{F_{\rm sy}}{F_{\rm IC}}
\varpropto B^2$, we expect that for sufficiently high magnetic fields,
we will reach the regime where the Compton flux from the cascades is
equal to or smaller than the synchrotron flux in which case our numerical
scheme is no longer applicable.

\section{\label{3C279}The Big Blue Bump}

The spectral Energy distribution of AGN in the ultraviolet (UV) to soft X-ray
band ($ \sim 10$~eV-$1$~keV) is notoriously difficult to observe because of
dust and gas in our galaxy and the AGN environment. The SEDs of many blazars
exhibits a UV soft X-ray excess, called the big blue bump (BBB)
\citep[]{pian99,palma11,raiteri05,raiteri06,raiteri07}. It is often
attributed to the thermal emission from the accretion disk. In blazars,
its signature is often particularly hard to detect because of dominant
non-thermal emission from the jet. Understanding the origin of the BBB
is important since this provides information on the central engine of
the AGN.

3C~279 was among the first blazars discovered as a $\gamma$-ray source with
the Compton Gamma-Ray Observatory \citep[]{hartman92}. In 2007 it was detected
as a VHE $\gamma$-ray source with the MAGIC I telescope, making it the most
distant known VHE $\gamma$-ray source at a redshift of $0.536$ \citep[]{HB93}.
Its relativistic jet is oriented at a small angle to the line of the sight
of $< 0.5^0$ \citep[]{J04}. It is also detected by \emph{Fermi} \citep[]{abdo09c}
with photon spectral index $2.23$. There is evidence of a spectral break of
around a few GeV to a photon spectral index of $2.50$. It is strongly believed
that the radio to optical emission is due to synchrotron radiation by relativistic
particles in the jet. However, the origin of the high energy emission is still not
well understood \citep[see, e.g.,][]{br09}.

\cite{pian99} monitored 3C~279 in the ultraviolet, using IUE, and combined
their data with higher-energy observations from ROSAT and EGRET from 1992 December
to 1993 January. During this period, the source was in a very low state, allowing
for the detection of a UV excess (the BBB), which is typically hidden below a
dominant power-law continuum attributed to non-thermal emission from the jet.
\cite{pian99} proposed that the $\gamma$-ray emission in the SED of 3C~279
is produced by the external Compton mechanism, and suggested that the observed
UV excess might be due to thermal emission from an accretion disk.

As an alternative to thermal emission from the accretion disk, \cite{S97} proposed
the bulk Compton mechanism as a possible explanation of a UV/X-ray excess in quasar
SEDs. If the jet contains a substantial population of cold (i.e., thermal,
non-relativistic or mildly relativistic) electrons, they could scatter
external optical/UV photons with the bulk Lorentz factor of $\Gamma \thicksim 10$,
resulting in bulk Compton radiation in the far UV or soft X-ray range.

\begin{figure}[ht]
\vskip 1.5cm
\centering
\includegraphics[width=10cm]{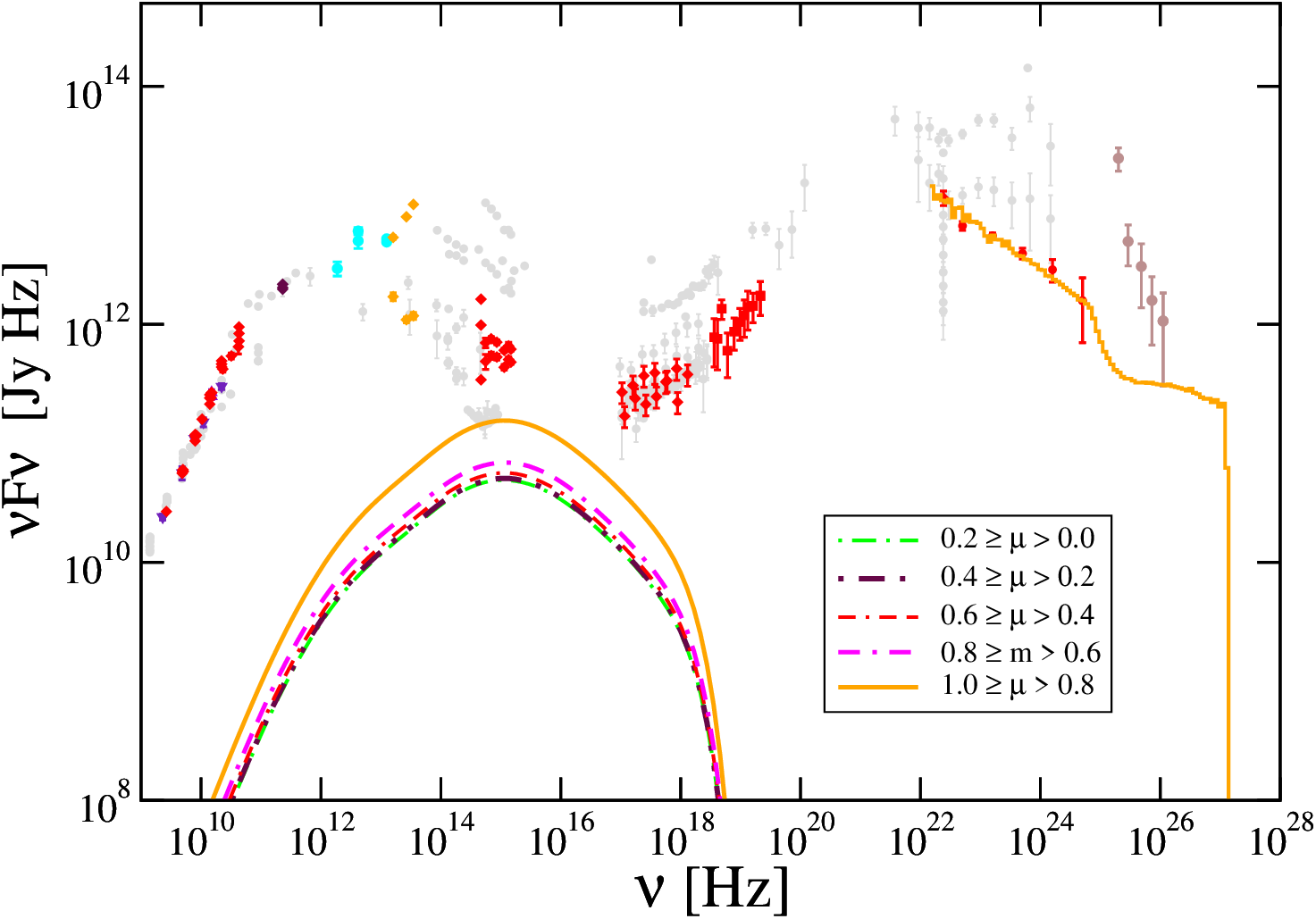}
\caption{\label{fit3C279}Illustration of a possible BBB in 3C~279 from
cascade synchrotron emission.
Parameters: $B= 10^{-2}$~G, $\theta_B = 85^0$;
 $R_{\rm ext} = 5\times10^{17}$~cm, $T = 2000$~K,
$\alpha = 2.37$, $u_{\rm ext} = 10^{-4}$~erg~cm$^{-3}$, {$E_{\gamma, {\rm max}}
= 5$~TeV}. Data from \cite{abdo10}.
}
\end{figure}

Here we suggest an alternative contribution to the BBB feature from
cascade synchrotron emission. Figures 2 -- 4
illustrate that the synchrotron emission from cascades may peak in the UV/X-ray
range, thus mimicking a BBB for sufficiently strong magnetic fields
($B \gtrsim 1$~mG). Figure \ref{fit3C279} illustrates 
the contribution that synchrotron emission from VHE $\gamma$-ray
induced pair cascades can make to the BBB in 3C~279.
The primary HE $ \gamma$-ray spectrum with a photon spectral index of
$\alpha = 2.5$ matcheds the Fermi spectrum of 3C~279.
The external radiation field is parameterized through $u_{\rm ext} =
10^{-4}$~erg~cm$^{-3}$ and $R_{\rm ext} = 5\times10^{17}$~cm,
and the parameters combine to the luminosity of $L = 4 \pi R_{\rm ext}^2 \, c
\, u_{\rm ext}
\sim 10^{43}$~erg~s$^{-1}$ corresponding to a $\nu F_{\nu}$ peak
flux of $\thicksim 10^9$~JyHz, about $2$ orders of magnitude below the observed
IR/optical -- UV flux level. The magnetic field is $B = 10^{-2}$~G, oriented at
an angle of $\theta_B = 85^o$. The incident $\gamma$-ray spectrum extends out
to $E_{\gamma, {\rm max}} = 5$~TeV, and the external radiation field is modeled
as a blackbody with a temperature of $ T= 2000$~K (corresponding to a peak of the
blackbody spectrum at a photon energy of $E_s^{\rm pk}= 0.5$~eV). This leads
to a $\gamma\gamma$ absorption cut-off at an energy $E_c = (m_e c^2)^2 / E_s
\sim 1$~TeV.

We suggest that synchrotron emission from VHE $\gamma$-ray induced pair
cascades can enhance the BBB feature in the SEDs of several
blazars such as 3C~279. An observational test of this hypothesis may be
provided through spectropolarimetry. A BBB due to (unpolarized) thermal
emission from an accretion disk will produce a decreasing percentage of
polarization with increasing frequency throughout the optical/UV range.
In contrast, if the BBB is produced as synchrotron emission from cascade
pairs in globally ordered magnetic fields, it is also expected to be polarized.
Therefore, we predict that a BBB due to cascade synchrotron emission would result
in a degree of polarization showing only a weak dependence on frequency over the
optical/UV range. As an example, in recent observations of the high-redshift
$\gamma$-ray loud quasar PKS~0528+134, \cite{palma11} found a decreasing
degree of polarization with increasing frequency throughout the optical range,
arguing for an increasing contribution from thermal emission towards the blue
end of the optical spectrum.

\section{\label{summary}Summary}

We investigated the magnetic-field dependence and synchrotron emission
signatures of Compton-supported pair cascades initiated by the interaction
of nuclear VHE $\gamma$-rays with arbitrary external radiation fields, 
for a model-independent, generic power-law shape of the primary
VHE $\gamma$-ray emission.
We
follow the spatial development of the cascade in full 3-dimensional geometry
and study the dependence of the radiative
output on various parameters pertaining to the external radiation field and
the magnetic field in the cascade region.
We confirm that synchrotron radiation from the cascades is negligible in
NGC~1275 and Cen~A for the parameters we used in our previous works. We
demonstrated that the magnetic field can not be well constrained by considering
the high-energy (Compton) output from the cascade emission alone, without
observational signatures from their synchrotron emission. This was illustrated
for the case of NGC~1275, for which we could produce equally acceptable fits
to the Fermi spectrum for a variety of magnetic-field values, which resulted
in substantially different synchrotron signatures.

We have shown that synchrotron emission from VHE $\gamma$-ray induced pair
cascades may produce UV/X-ray signatures resembling the BBB observed in the
SEDs of several blazars, in particular in their low states. 
We used the example of 3C~279 to illustrate that cascade synchrotron
emission may make a substantial contribution to the BBB feature.
We point out that spectropolarimetry may serve
as a possible observational test to distinguish a thermal from a non-thermal
(cascade) origin of the BBB.

\acknowledgements{ This work was supported by NASA through Fermi Guest
Investigator Grants NNX09AT81G and NNX10AO49G. We thank the anonymous 
referee for valuable suggestions. }

\end{document}